\documentclass[%
 reprint,
 amsmath,amssymb,
 aps,
]{revtex4-2}

\usepackage{graphicx} 
\usepackage{dcolumn}  
\usepackage{bm}       
\usepackage{amsmath}
\usepackage{microtype}

\usepackage[colorlinks=true,
            linkcolor=blue,
            citecolor=blue,
            urlcolor=blue]{hyperref}

\begin{document}

\preprint{APS/123-QED}

\title{Phase-Selective Excitation of Betatron Oscillations by Nonadiabatic Magnetic-Field Switching }

\author{R.S. Anandu}
 
\author{B. Ramakrishna}%
 \email{bhuvan@phy.iith.ac.in}
\affiliation{%
 Department of Physics, Indian Institute of Technology Hyderabad, Kandi, Sangareddy, 502284, India.\\}%

\date{\today}

\begin{abstract}
Nonadiabatic removal of an external transverse magnetic field provides a phase-selective mechanism for controlling betatron oscillations in laser wakefield accelerators. When the field is switched off on a timescale shorter than the betatron period, the equilibrium orbit shifts abruptly and acts as an impulsive transverse drive. The induced motion interferes coherently with the preexisting betatron oscillation, leading to phase-dependent enhancement or suppression of the oscillation amplitude. A theoretical model shows that the excitation is governed by the dimensionless switching parameter $\chi=\omega_\beta L_s/c$, which distinguishes nonadiabatic and adiabatic regimes. Particle-in-cell simulations confirm the predicted scaling and demonstrate controllable modulation of the betatron radiation spectrum while leaving longitudinal acceleration largely unaffected. These results establish magnetic-field switching as a direct mechanism for phase control of relativistic betatron oscillations in plasma-based accelerators.

\end{abstract}

\maketitle

\section{Introduction}

Laser–wakefield accelerators (LWFAs) provide compact sources of relativistic
electron beams with accelerating gradients exceeding ${\rm GV\,m^{-1}}$
\cite{Esarey2009}. The first demonstrations of quasi-monoenergetic beams
\cite{Faure2004,Geddes2004} established the nonlinear blowout regime, in which
electrons undergo transverse betatron oscillations in the focusing field of
the ion cavity and emit broadband femtosecond X rays with synchrotron-like
characteristics \cite{Corde2013,Albert2016}. Bright spatially coherent betatron radiation
and its spectral properties were experimentally demonstrated
\cite{Kneip2010,Fourmaux2011}, enabling applications such as phase-contrast
imaging and microtomography \cite{Kneip2011,Wenz2015}. 

The spectrum and brightness of betatron radiation depend strongly on the oscillation amplitude of accelerated electrons, making control of transverse electron motion critically important. Coherent laser-driven betatron oscillations were reported by Nemeth \emph{et al.}~\cite{Nemeth2008}. Subsequent studies demonstrated control of the oscillation amplitude and phase through longitudinal density tailoring~\cite{kozlova2020hard,rakowski2022transverse}. Resonant enhancement of betatron emission via controlled injection dynamics was reported in~\cite{huang2016resonantly}, while asymmetric laser pulses have been used to further enhance radiation~\cite{ferri2018enhancement}. In these approaches, transverse dynamics are modified indirectly through density shaping, injection engineering, or self-consistent wake evolution. External magnetic fields have also been explored in laser wakefield acceleration to influence injection, guiding, and wake structure~\cite{rassou2015influence,vieira2012magnetically,Vieira2011,hosokai2006effect}. However, the direct excitation of betatron motion through time-dependent magnetic-field variation remains largely unexplored.

\begin{figure}[h]
 \centering
\includegraphics[width=230pt]{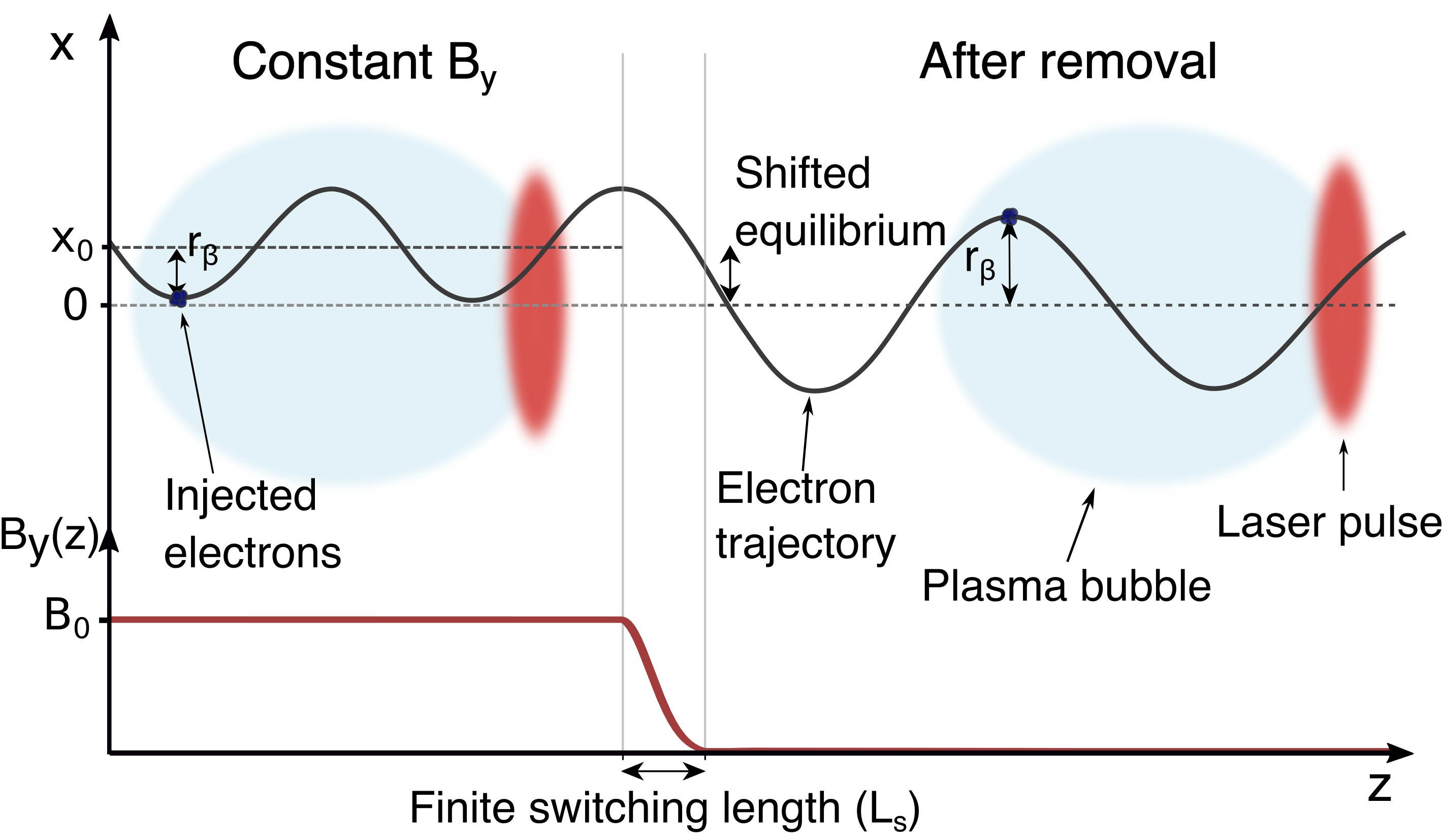}
\caption{\label{fig.1}
Schematic of phase-dependent excitation of betatron oscillations caused by switching an external transverse magnetic field over a finite length $L_s$.
}
\end{figure}

In this work, we introduce a mechanism for phase-selective manipulation of betatron oscillations based on controlled switching of an externally applied transverse magnetic field. The dynamics are governed by the ratio between the magnetic-field switching time and the intrinsic betatron period. When the field varies slowly compared with the oscillation period (adiabatic limit), electrons follow the evolving equilibrium orbit and no additional transverse excitation occurs. In contrast, when the magnetic field is removed on a timescale comparable to or shorter than the betatron period, the switching induces an additional transverse oscillation, as illustrated in Fig.~\ref{fig.1}. This induced motion coherently superposes with the pre-existing betatron oscillation. The phase at the end of the switching region determines whether the interference is constructive or destructive, leading to enhancement or suppression of the final oscillation amplitude. We show that nonadiabatic magnetic-field switching provides a direct phase-control mechanism for relativistic betatron oscillations, enabling tunable manipulation of betatron radiation without altering the wakefield structure.

In the nonlinear blowout regime of laser wakefield acceleration, relativistic electrons propagate inside an ion cavity that provides a linear transverse focusing force \cite{Esarey2009},
\begin{equation}
F_x = -\frac{m_e \omega_p^2}{2}\, x ,
\end{equation}
where $\omega_p=\sqrt{n_e e^2/(m_e\varepsilon_0)}$. The transverse motion satisfies
\begin{equation}
\ddot{x}+\omega_\beta^2 x=0,
\label{eq:betatron}
\end{equation}
with $\omega_\beta=\omega_p/\sqrt{2\gamma}$. 
The solution $x(t)=r_\beta\cos(\omega_\beta t+\phi)$ describes betatron oscillations of amplitude $r_\beta$. 
Over the switching interval considered below, the variation of $\gamma$ due to longitudinal acceleration is assumed negligible, so that $\omega_\beta$ can be treated as approximately constant.

We now introduce an external transverse magnetic field
\begin{equation}
\mathbf{B}(z)=B_y(z)\,\hat{y}.
\end{equation}
For $v_z\simeq c$, the transverse equation becomes
\begin{equation}
\ddot{x}+\omega_\beta^2 x
=
\frac{e c}{\gamma m_e} B_y(z).
\label{eq:forced}
\end{equation}
For constant $B_y=B_0$, the motion occurs about a shifted equilibrium,
\begin{equation}
x_0=\frac{e c}{\gamma m_e \omega_\beta^2}B_0
=\frac{2 e c}{m_e\omega_p^2}B_0,
\label{eq:eqshift}
\end{equation}
which is independent of $\gamma$.

When $B_y$ varies along $z$, the equilibrium position becomes time dependent. Writing $x=x_0(t)+\xi(t)$ gives
\begin{equation}
\ddot{\xi}+\omega_\beta^2\xi=-\ddot{x}_0(t),
\label{eq:driven}
\end{equation}
so that changes in the equilibrium orbit act as a driving term. 
Assuming $\xi(-\infty)=\dot{\xi}(-\infty)=0$, the induced betatron amplitude is
\begin{equation}
\Delta r_\beta
=
\frac{1}{\omega_\beta}
\left|
\int_{-\infty}^{\infty}
\ddot{x}_0(t)\,
e^{i\omega_\beta t}\,dt
\right|.
\label{eq:general_amp}
\end{equation}
The excitation is determined by the spectral content of the equilibrium variation at $\omega_\beta$, so that the magnetic-field switching acts as a spectral filter whose response depends on the frequency content of the chosen profile.

\paragraph{Switching-length scaling.}
Let $L_s$ denote the characteristic switching length and define $\chi=\omega_\beta L_s/c$. Then
\begin{equation}
\Delta r_\beta = x_{0,\max} F(\chi),
\end{equation}
with $\Delta r_\beta\rightarrow x_{0,\max}$ for $\chi\ll1$ (impulsive limit) and $\Delta r_\beta\rightarrow0$ for $\chi\gg1$ (adiabatic limit). 
For exponential switching, $F(\chi)=(1+\chi^2)^{-1/2}$, 
while for the Gaussian profile employed in the simulations, 
$F(\chi)\sim\exp(-\chi^2/2)$ (see Supplemental Material for derivations).

\paragraph{Phase-dependent response.}
Electrons entering the switching region undergo betatron motion 
$x(t)=r_\beta\cos(\omega_\beta t+\phi)$.
Let $\theta=\omega_\beta t_{\mathrm{sw}}+\phi$ denote the betatron phase at the onset of magnetic-field removal.
During switching, the phase advances by 
$\Delta\theta \simeq \omega_\beta \tau_s \equiv \chi$,
where $\tau_s \simeq L_s/c$ is the effective switching time.
The interference phase after switching is therefore
$\theta_f=\theta+\chi$.

The field removal induces an additional transverse oscillation that coherently superposes with the pre-existing motion, yielding the final amplitude
\begin{equation}
r_{\beta,\mathrm{final}}
=
\sqrt{
r_\beta^2
+
\Delta r_\beta^2
+
2 r_\beta \Delta r_\beta \cos\theta_f
}.
\label{eq:final_amp}
\end{equation}

\begin{figure}[h]
 \centering
\includegraphics[width=250pt]{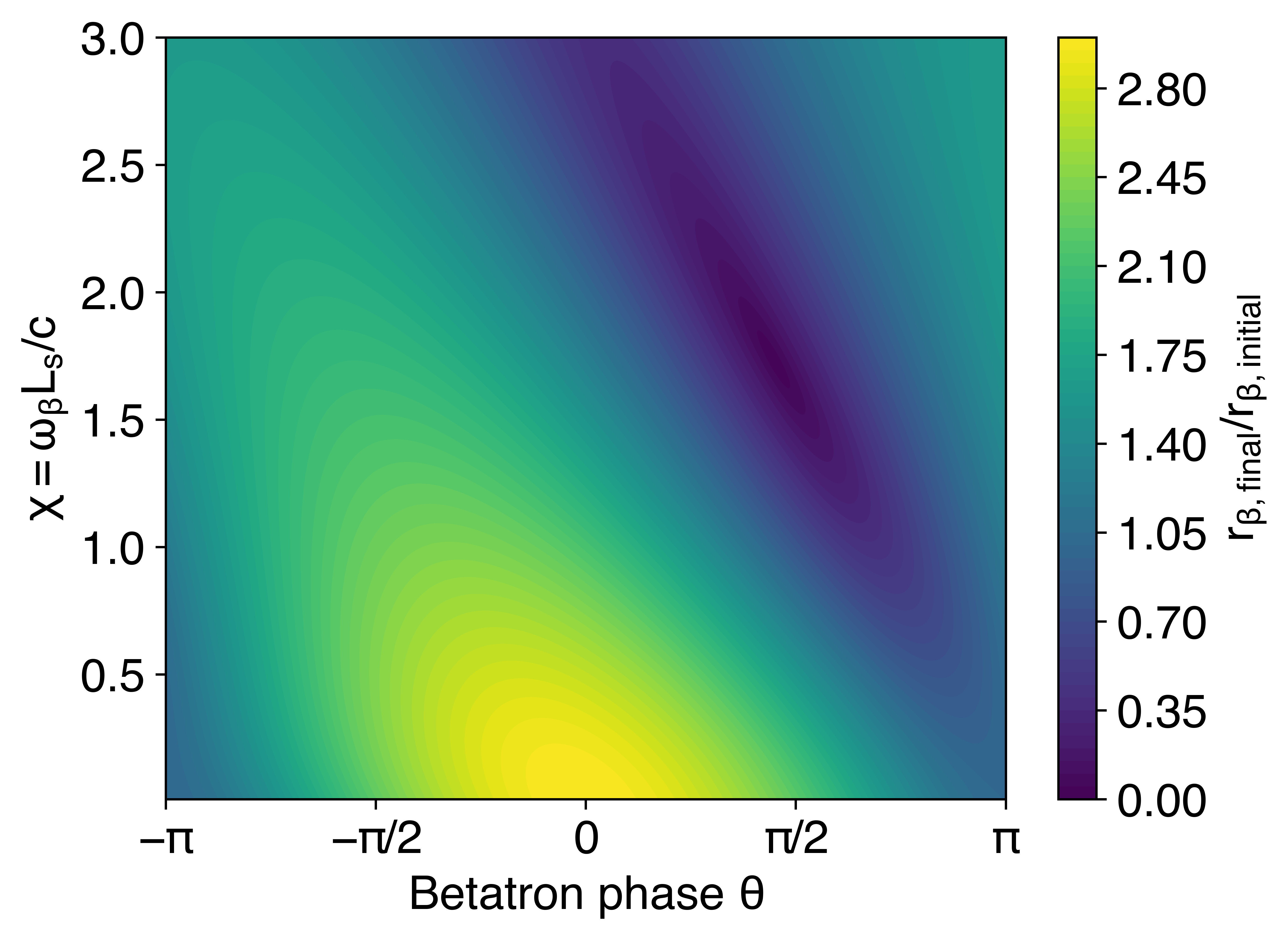}
\caption{\label{fig.2}
Normalized final betatron amplitude 
$r_{\beta,\mathrm{final}}/r_{\beta,\mathrm{initial}}$ as a function of the initial phase $\theta$ and the switching parameter $\chi=\omega_\beta L_s/c$ for $x_{0,\max}=1.5\,r_{\beta,\mathrm{initial}}$, assuming a Gaussian switching profile. 
Constructive ($\theta_f=0$) and destructive ($\theta_f=\pm\pi$) interference between the induced and initial oscillations lead to enhancement and suppression of the final amplitude. 
The phase modulation is strongest for $\chi\lesssim1$ and weakens as the switching duration increases toward the adiabatic regime.
}
\end{figure}

Equation~(\ref{eq:final_amp}) reveals three regimes determined by the ratio
$\alpha=\Delta r_\beta/r_\beta$ between the switching-induced oscillation and the
initial betatron amplitude. For $\alpha<1$, corresponding to electrons with initially large betatron amplitudes, the induced motion produces only partial modulation of the oscillation. When $\alpha=1$, the induced oscillation equals the initial
amplitude, so that destructive interference ($\theta_f=\pi$) can cancel the
betatron motion. For $\alpha>1$, corresponding to initially small oscillations,
the induced motion dominates and can strongly enhance the final amplitude.
The resulting interference map in $(\theta,\chi)$ space is shown in
Fig.~\ref{fig.2}. Nonadiabatic switching therefore provides a direct mechanism
for phase-selective control of betatron oscillations and their associated
radiation.

\begin{figure*}[t]
\centering
\includegraphics[width=\textwidth]{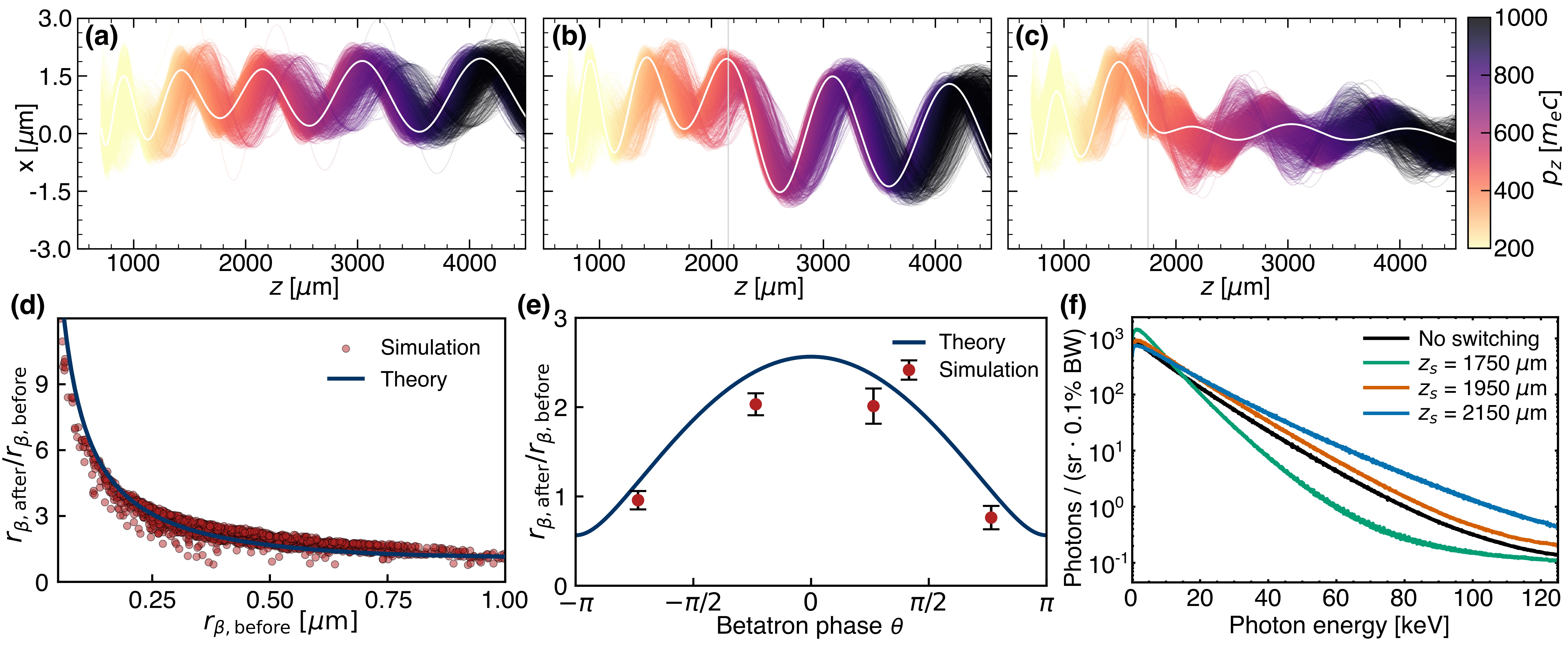}
\caption{ \label{fig:position}
Phase-dependent modification of betatron oscillations by nonadiabatic magnetic-field switching.
(a)–(c) Transverse electron position $x$ as a function of propagation distance $z$ along the accelerator, showing trajectories for no switching and for switching at $z_s=2150 \mu m$ and $1750 \mu$m with switching length $L_s=90 \mu$m respectively.
(d) Final versus initial betatron amplitude showing agreement between particle-in-cell simulations (markers) and theory (line).
(e) Normalized final amplitude $r_{\beta,\mathrm{after}}/r_{\beta,\mathrm{before}}$ as a function of betatron phase $\theta$ at the switching location.
(f) Corresponding betatron radiation spectra for different switching positions compared with the no-switching case.
}
\end{figure*}


Particle-in-cell (PIC) simulations are performed using the quasi-cylindrical framework \textsc{fbpic}~\cite{Lehe2016}. 
The domain extends $60\,\mu\mathrm{m}$ longitudinally and $25\,\mu\mathrm{m}$ radially, 
with resolutions $\Delta z = 40\,\mathrm{nm}$ and $\Delta r = 100\,\mathrm{nm}$, 
using $N_m=4$ azimuthal modes and a spectral solver of order $n_{\mathrm{order}}=32$ 
in a relativistic moving window. 
The plasma is initialized with $2\times2\times16$ macroparticles per cell. 
A laser pulse ($a_0=2.2$, $\lambda_0=0.8~\mu$m) drives a nonlinear wake in plasma of density 
$n_e=1.7\times10^{18}~\mathrm{cm^{-3}}$. 
Electron injection is triggered by a sharp plasma-density transition. 
An external transverse magnetic field with peak amplitude $50~\mathrm{T}$ 
is applied and removed with a Gaussian profile, with switching time and position varied to probe the nonadiabatic regime. 
Electron trajectories are tracked and the resulting betatron radiation is computed using \textsc{SynchRad}~\cite{Andriyash2018}.


\begin{figure*} 
\centering
\includegraphics[width=\textwidth]{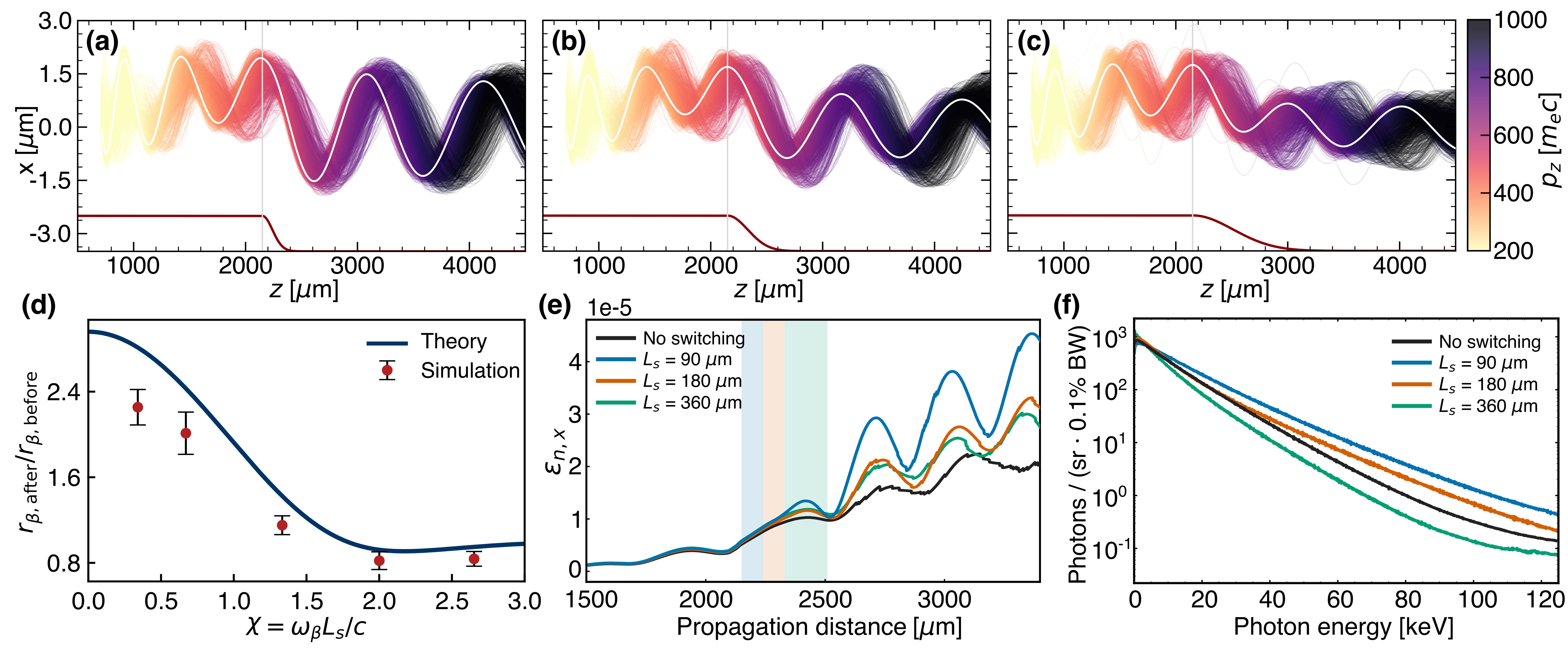}
\caption{\label{fig:duration}
Transition from the impulsive nonadiabatic to the adiabatic switching regime with increasing magnetic-field switching length $L_s$.
(a)–(c) Transverse trajectories $x(z)$ of tracked high-energy electrons ($\gamma>500$) for switching lengths $L_s=90~\mu$m, $180~\mu$m, and $360~\mu$m, respectively; the red curve indicates the applied magnetic-field profile along the propagation direction.
(d) Amplification of the betatron oscillation, defined as $r_{\beta,\mathrm{after}}/r_{\beta,\mathrm{before}}$, as a function of the switching parameter $\chi=\omega_\beta L_s/c$ for electrons with initial radius $r_\beta\approx0.5~\mu$m and initial phase $\theta\approx0$; simulation results (markers) are compared with the analytic prediction (line).
(e) Evolution of the normalized transverse emittance $\varepsilon_{n,x}$ for different switching lengths, where the shaded regions indicate the spatial extent of magnetic-field switching for each case.
(f) Corresponding betatron radiation spectra for $L_s=90$, $180$, and $360~\mu$m compared with the no-switching case, showing reduced high-energy photon yield as the switching becomes slower.
}

\end{figure*}
%
Figure~\ref{fig:position} illustrates the phase-selective excitation of betatron oscillations in the near-impulsive regime, where the magnetic field is removed over a short length $L_s=90~\mu\mathrm{m}$ ($\chi=\omega_\beta L_s/c\lesssim1$), inducing an additional transverse oscillation $\Delta r_\beta$ that coherently interferes with the pre-existing betatron motion and leads to enhancement or suppression of the final oscillation amplitude depending on the betatron phase.

Figure~\ref{fig:position}(a)–(c) show transverse electron trajectories $x(z)$ for three cases: no magnetic-field switching, switching at $z_s=1750~\mu\mathrm{m}$, and switching at $z_s=2150~\mu\mathrm{m}$. The switching locations in cases (b) and (c) correspond approximately to betatron phases $\theta\approx0$ and $\theta\approx-\pi$, respectively. In the presence of the external magnetic field, electrons oscillate around a shifted equilibrium position, as illustrated in Fig.~\ref{fig:position}(a); for the $50~\mathrm{T}$ field used here this equilibrium shift is $x_0\approx0.98~\mu\mathrm{m}$. When the field is removed near $\theta\approx0$ [Fig.~\ref{fig:position}(b)], the induced oscillation interferes constructively with the preexisting betatron motion, leading to a strong enhancement of the oscillation amplitude. In contrast, switching near $\theta\approx-\pi$ [Fig.~\ref{fig:position}(c)] produces destructive interference and reduction of the transverse amplitude. 

The amplification of the oscillation radius is quantified in Fig.~\ref{fig:position}(d). From Eq.~(\ref{eq:final_amp}), the enhancement of the final amplitude is governed by the ratio $\alpha=\Delta r_\beta/r_\beta$, which measures the strength of the switching-induced oscillation relative to the initial betatron motion. Electrons with smaller initial oscillation radii correspond to larger values of $\alpha$ and therefore experience stronger amplification, whereas electrons with larger initial amplitudes exhibit weaker enhancement. Figure~\ref{fig:position}(d) shows the amplification factor $r_{\beta,\mathrm{after}}/r_{\beta,\mathrm{before}}$ as a function of the initial betatron radius $r_{\beta,\mathrm{before}}$. Electrons with small initial radii are amplified most strongly, reaching final amplitudes approximately $6$–$12$ times larger than their initial values, while electrons with larger initial radii show only modest enhancement. The simulation results agree closely with the theoretical prediction of Eq.~(\ref{eq:final_amp}). In this comparison we select electrons that undergo switching near the constructive-interference phase ($\theta\approx0$), where the amplification is maximal.

The phase dependence of the excitation is shown in Fig.~\ref{fig:position}(e), where the normalized final amplitude $r_{\beta,\mathrm{after}}/r_{\beta,\mathrm{before}}$ is plotted as a function of the betatron phase $\theta$. To isolate the phase dependence, electrons with similar initial oscillation radii ($r_\beta\approx0.5~\mu\mathrm{m}$) are selected. The analytic model predicts that the excitation arises from interference between the initial betatron oscillation and the switching-induced transverse impulse. Because the switching occurs over a finite duration, the betatron phase advances during the switching interval by $\Delta\theta\approx\chi$, so that the effective interference phase after switching is shifted relative to the initial phase shown on the horizontal axis. After accounting for this phase advance, the simulation results follow the predicted phase dependence and reproduce the overall shape of the theoretical curve. Small quantitative deviations arise because the effective switching duration experienced by the electrons is not exactly equal to the Gaussian scale parameter $\sigma$, and because the betatron frequency evolves gradually during propagation as the electron energy increases. Despite these effects, the simulations clearly capture the predicted phase dependence of the excitation.

Fig.~\ref{fig:position}(f) shows the corresponding betatron radiation spectra
for the different switching cases. Consistent with the trajectory dynamics,
switching near the constructive phase leads to a pronounced enhancement of the
high-energy photon yield, whereas switching near the destructive phase
suppresses the radiation relative to the no-switching case. The enhancement is
most significant in the hard X-ray region, while the low-energy part of the
spectrum remains comparatively less affected due to the distribution of
betatron phases within the electron ensemble. These results demonstrate that
rapid magnetic-field removal provides a direct mechanism for phase-selective
excitation of betatron oscillations and enables tunable control of the emitted
betatron radiation spectrum. Despite this phase spread, the simulations show that rapid switching produces a clear net modulation of the radiation spectrum.

The transition from impulsive to finite-duration switching is examined by varying the magnetic-field switching length $L_s$, which determines the parameter $\chi=\omega_\beta L_s/c$. Increasing $\chi$ means that the magnetic-field removal occurs over a larger fraction of the betatron period, allowing the betatron phase to evolve during the interaction and progressively reducing the coherence of the switching-induced excitation.

Figure~\ref{fig:duration}(a)–(c) show the transverse trajectories $x(z)$ of tracked high-energy electrons ($\gamma>500$) for switching lengths $L_s=90~\mu$m, $180~\mu$m, and $360~\mu$m, respectively. The red curve indicates the applied magnetic-field profile along the propagation direction. For the shortest switching length ($L_s=90~\mu$m), the rapid field removal acts as an impulsive transverse kick and produces a strong enhancement of the betatron oscillation. As the switching duration increases, the excitation weakens and the trajectories progressively approach those obtained in the absence of switching. This trend reflects the loss of coherent phase interference as the betatron phase evolves during the switching process.

The dependence of the excitation on the switching parameter is quantified in Fig.~\ref{fig:duration}(d), which shows the amplification factor $r_{\beta,\mathrm{after}}/r_{\beta,\mathrm{before}}$ as a function of $\chi$. In the small-$\chi$ limit the amplification approaches the impulsive prediction of the analytic model, confirming that rapid magnetic-field removal acts as an effective transverse impulse. As $\chi$ increases, the amplification decreases systematically because the switching-induced drive is distributed over a longer interaction time, approaching the adiabatic regime. 

The simulation results reproduce the predicted scaling with $\chi$ and capture the overall transition from impulsive to adiabatic switching. Small quantitative deviations from the analytic curve arise from the simplifying assumptions of the model. In particular, the analytic treatment assumes a constant betatron frequency and an idealized switching profile characterized by a single scale length, whereas in the simulations the betatron frequency $\omega_\beta$ evolves gradually as the electron energy increases and the magnetic-field switching occurs over a finite spatial extent. Despite these effects, the simulations follow the theoretical trend closely over the entire parameter range, confirming the validity of the analytic description of the switching dynamics.

The impact of the switching duration on the beam phase-space dynamics is shown in Fig.~\ref{fig:duration}(e), which presents the evolution of the normalized transverse emittance $\varepsilon_{n,x}$ for different switching lengths. Rapid switching produces an impulsive transverse kick that drives large betatron oscillations and leads to significant phase-space heating, resulting in larger emittance growth. As the switching becomes more gradual, electrons follow the evolving equilibrium orbit more adiabatically and the transverse phase-space distortion is reduced. The shaded regions indicate the spatial extent of the magnetic-field switching for each case, illustrating how the excitation is distributed over the interaction region.

Fig.~\ref{fig:duration}(f) shows the corresponding betatron radiation spectra for switching performed at the same location $z_s=2150~\mu\mathrm{m}$ with different switching durations. For rapid switching the betatron phase evolves only weakly during the interaction, so the induced oscillation remains near the constructive phase and enhances the high-energy photon yield. As the switching duration increases, the phase advances during the switching interval, causing electrons to sample both constructive and destructive interference conditions and reducing the net excitation. Consequently the radiation spectrum progressively approaches the no-switching case. These results demonstrate that the magnetic-field switching length provides a direct control parameter for tuning both the transverse electron dynamics and the resulting betatron radiation.

In conclusion, we have demonstrated that nonadiabatic removal of an externally applied transverse magnetic field provides a mechanism for phase-selective excitation of betatron oscillations in laser wakefield accelerators. When the magnetic field is switched off on a length scale short compared with the betatron wavelength, the resulting shift of the equilibrium orbit induces an additional transverse oscillation that coherently interferes with the pre-existing betatron motion. The final oscillation amplitude is therefore determined by the interference phase and by the ratio $\alpha=\Delta r_\beta/r_\beta$, enabling either enhancement or suppression of the transverse oscillation. A theoretical model shows that the excitation amplitude is determined by the Fourier component of the equilibrium-orbit motion at the betatron frequency. Magnetic-field switching therefore acts as a spectral filter characterized by the parameter $\chi=\omega_\beta L_s/c$. Particle-in-cell simulations confirm the predicted phase-dependent excitation, the scaling with switching length, and the resulting modulation of the betatron radiation spectrum while leaving longitudinal acceleration largely unaffected.

The required magnetic fields and switching scales are within reach of present high-field laser–plasma technologies. Laser-driven capacitor–coil targets have demonstrated magnetic fields in the kilotesla range over submillimeter spatial scales \cite{Fujioka2013,Zhang2018}, while pulsed microcoil systems are capable of generating tens of tesla fields in compact geometries \cite{Mackay2000}. Such approaches provide feasible routes for producing controlled magnetic-field profiles in laser–plasma experiments, enabling implementation of the switching mechanism proposed here.

These results establish nonadiabatic magnetic-field switching as a new mechanism for externally controlling relativistic transverse dynamics in plasma accelerators, providing a pathway toward tunable betatron radiation sources and new methods for manipulating beam phase space.

\begin{acknowledgments}
The authors gratefully acknowledge the National Supercomputing Mission (NSM) for providing access to the \textit{PARAM Seva} computing facility at IIT Hyderabad. NSM is implemented by C-DAC and supported by the Ministry of Electronics and Information Technology (MeitY) and the Department of Science and Technology (DST), Government of India. This work was also supported by the Ministry of Education under the STARS program (Project No.\ MoE\_STARS-2/2023-0233), funded by the Science and Engineering Research Board (SERB), Government of India. R.\ S.\ Anandu acknowledges the Council of Scientific and Industrial Research (CSIR), Government of India, for the research fellowship.
\end{acknowledgments}

\bibliography{apssamp}


\clearpage
\onecolumngrid
\setcounter{secnumdepth}{0}

\end{document}